\documentclass{iopart}
\usepackage{iopams}
\usepackage{graphicx}

\def\EW#1{{%
  \ifmmode <\!#1\!>\else $<\!#1\!>$\fi}}
\def\ket#1{{%
  \ifmmode |\,#1\,\rangle \else $|\,#1\,\rangle$\fi}}
\def\bra#1{{%
  \ifmmode \langle\,#1\,| \else $\langle\,#1\,|$\fi}}
\def\Ket#1{{%
  \ifmmode \left|\,#1\,\right\rangle \else $\left|\,#1\,\right\rangle$\fi}}
\def\Bra#1{{%
  \ifmmode \left\langle\,#1\,\right| \else $\left\langle\,#1\,\right|$\fi}}
\def\braket#1#2{{%
  \ifmmode \langle\,#1\,|\,#2\,\rangle \else
  $\langle\,#1\,|\,#2\,\rangle$ \fi}}
\def\braCket#1#2#3{{%
  \ifmmode \langle\,#1\,|\,#2\,|\,#3\,\rangle \else
  $\langle\,#1\,|\,#2\,|\,#3\,\rangle$ \fi}}

\makeatletter
\def\Braket#1#2{%
 \setbox\@tempboxa\hbox{$#1$} \@tempskipa\wd\@tempboxa
 \setbox\@tempboxa\hbox{$#2$}
 \ifdim\ht\@tempboxa<\@tempskipa
   \left\langle\left.\,#1\,\right|\,#2\,\right\rangle
 \else
   \left\langle\,#1\,\left|\,#2\,\right.\right\rangle
 \fi
}
\def\BraCket#1#2#3{  
 \setbox\@tempboxa\hbox{$#1$} \@tempskipa\wd\@tempboxa
 \setbox\@tempboxa\hbox{$#2$} \@tempskipb\wd\@tempboxa
 \setbox\@tempboxa\hbox{$#3$}
 \ifdim\ht\@tempboxa>\@tempskipb
   \ifdim\ht\@tempboxa>\@tempskipa    
     \left\langle\,#1\,\left|\,#2\,\left|\,#3\,\right.\right.\right\rangle
   \else  
     \left\langle\left.\left.\,#1\,\right|\,#2\,\right|\,#3\,\right\rangle
   \fi
 \else
   \ifdim\@tempskipb>\@tempskipa  
     \left\langle\,#1\,\left|\left.\,#2\,\right|\,#3\,\right.\right\rangle
   \else 
     \left\langle\left.\left.\,#1\,\right|\,#2\,\right|\,#3\,\right\rangle
   \fi
 \fi
}
\makeatother

\begin{document}

\jl{9}

\title[Ballistic expansion of a dipolar condensate]{Ballistic expansion of a dipolar condensate}

\author{Stefano Giovanazzi,$^1$ Axel G\"{o}rlitz,$^2$
and Tilman Pfau,$^2$}

\address{$^1$ Center for Theoretical Physics and College of Science,
Polish Academy of Sciences, Aleja Lotnik\'ow
32/46, 02-668 Warsaw,
Poland\\
$^2$ 5th Institute of Physics, University of
Stuttgart, D-70550 Stuttgart, Germany.  }

\ead{stevbolz@yahoo.it}

\begin{abstract}
We have studied the free expansion of a Bose
condensate in which both the  usual s-wave
contact interaction and the dipole-dipole
interaction contribute considerably to the
total interaction energy. We calculate
corrections due to dipolar forces to the
expansion of such a condensate after release
from a trap. In the Thomas-Fermi limit, we
find that the modifications are to lowest
order independent from the total number of
atoms.
\end{abstract}


\submitto{\JOB}

\section{Introduction}

In the atomic Bose-Einstein condensates (BECs)
realized so far, the atoms interact
essentially only at very short distance
(Van-der Waals interaction) compared to the
separation between atoms. The only relevant
parameter that characterizes the interaction
is the s-wave scattering length
\cite{stringari}. Recently dipole-dipole
interactions in atomic BECs have attracted
increasing interest
\cite{yi00,goral00,santos00,baranov,goral,meystre01,deMille,giova02,santos02,giova02b}.
These interactions would largely enrich the
variety of phenomena to be observed due to
their long-range and anisotropic character.

Atoms like rubidium and sodium which have been
successfully Bose-condensed, posses a magnetic
dipole moment, but it is very small. We focus
here on another atom as a possible candidate
for the realization of a dipolar condensate,
namely chromium. Chromium atoms have a larger
magnetic moment $m = 6 \mu_B$, where $\mu_B$
is the Bohr magneton, and interact with non
negligible dipole dipole forces. These
anisotropic forces compete with s-wave
scattering. If the s-wave scattering length
(unknown for chromium) is not too large, then
dipolar forces could be observed directly
through small but appreciable modifications in
the ground state density distribution or even
in the occurrence of instability.

We discuss in this paper the dipolar effects
on the free expansion of a dipolar BEC. In
many cases, trapped condensates are too small
to obtain clear images in absorption imaging
and therefore the condensate is released from
the trap to expand in order to obtain larger
dimension. Because the dipolar effects in
chromium are expected to be not too large, it
is interesting to maximize the dipolar effects
in ballistic expansion by tuning external
parameters like the trap anisotropy.

\section{Magnetic dipole interaction}

We consider a dipolar gas of neutral atoms
with magnetic moment $m$ oriented along the
magnetic field
\begin{eqnarray} {\bf B}(t)=
B\left[\cos\varphi \hat{\bf{z}} + \sin\varphi
\left(\cos(\Omega t)\hat{\bf{x}}+\sin(\Omega
t)\hat{\bf{y}}\right)\right].
\end{eqnarray}
The frequency $\Omega$ of the radial component
of the magnetic field is chosen such that
$\omega_{Larmor}\gg \Omega \gg \omega_{trap}$,
where $\omega_{trap}$ is any of the trap
frequencies. This corresponds to the
assumption that the atoms are not moving
during one rotation of the magnetic field and
the magnetic moments will follow adiabatically
the total magnetic field. The resulting long
range part of the interatomic interaction is
\cite{giova02b}
\begin{eqnarray}
U_{\mathrm{dd}}({\bf r}) &=& - \frac{\mu_0
m^2}{4 \pi} \left({3\,\cos^2\varphi -1\over
2}\right) \left({3\,\cos^2\theta -1}\over
{r^3}\right), \label{interactionaveraged}
\end{eqnarray}
where $\theta$ is the angle between the
interatomic separation vector ${\bf r}$ and
the $z$-axes. Here $m$ is the atomic magnetic
moment and $\mu_0$ is the magnetic
permeability of the vacuum. The angle
$\varphi$ serves as a "knob" to tune the
dipolar interaction in real time ($\varphi$
may have a low frequency time dependence).

\section{Thomas-Fermi mean-field analysis}

Our analysis is based on the zero-temperature
($T=0$) mean-field approach already used in
the context of dipolar Bose gases
\cite{yi00,goral00,santos00}. Such a
description can be accomplished through the
Gross-Pitaevskii equation \cite{stringari} for
the condensate order parameter $\Psi ({\bf
r},t)$ (normalized to the number of atoms $N$)
\begin{eqnarray}
i \hbar \frac{\partial \Psi }{\partial t} =
\left[-\frac{\hbar^2{\bf \Delta}}{2 M}
+\frac{M \omega_r^2}{2} (x^2+y^2)
 +\frac{M
\omega_z^2}{2}  z^2 + V_{mf} \right]\Psi ,
\label{ggp}
\end{eqnarray}
where $M$ is the atomic mass, $\omega_r$
($\omega_z$) the radial (longitudinal) trap
frequency and where the mean-field potential
$V_{mf}$ is given by
\begin{eqnarray}
V_{mf}({\bf r}) =
\frac{4\pi\hbar^2a}{M}|\Psi({\bf r})|^{2} +
\int d{\bf r}'\,|\Psi({\bf r}')|^{2}\,
U_{\mathrm{dd}}({\bf r}-{\bf r}')  ,
\end{eqnarray}
where $a$ is taken to be a positive s-wave
scattering length.

A measure of the strength of the dipole-dipole
interaction relative to the s-wave scattering
energy is given by the dimensionless quantity
\begin{eqnarray}
\varepsilon_{\mathrm{dd}} = \frac{\mu_0 m^2
M}{12 \pi \hbar^2 a}.\label{epsilon}
\end{eqnarray}
$\varepsilon_{\mathrm{dd}}$ is defined in such
a way that when it is larger than one
($\varepsilon_{\mathrm{dd}}>1$) a homogenous
dipolar condensate is unstable against
collapse in the Thomas-Fermi limit
\cite{goral} and in the absence of an external
potential.

For chromium the s-wave scattering length
$a_{\mathrm{Cr}}$ is not known. If we assume
that $a_{\mathrm{Cr}}$ is equal to the sodium
scattering length ($a=2.8$ nm) the dipolar
strength parameter becomes
$\varepsilon_{\mathrm{dd}}^{\mathrm{Cr}}=0.29$
($m=6\mu_B$, where $\mu_B$ is the Bohr
magneton).  In comparison, for rubidium and
sodium the corresponding values of
$\varepsilon_{\mathrm{dd}}$ are much smaller.
In contrast, for heteronuclear molecules, we
expect a much larger
$\varepsilon_{\mathrm{dd}}$.

The starting point for our analysis is the
limit of a large number of atoms, namely the
Thomas-Fermi limit in which the kinetic energy
has a very small effect on the ground state
density distribution. This physically relevant
case represents also an ideal situation from
an experimental point of view because an
observable quantity like the modification of
the condensate shape is \emph{independent of
the number of atoms}.

\begin{figure}
\includegraphics[width=0.7\columnwidth]{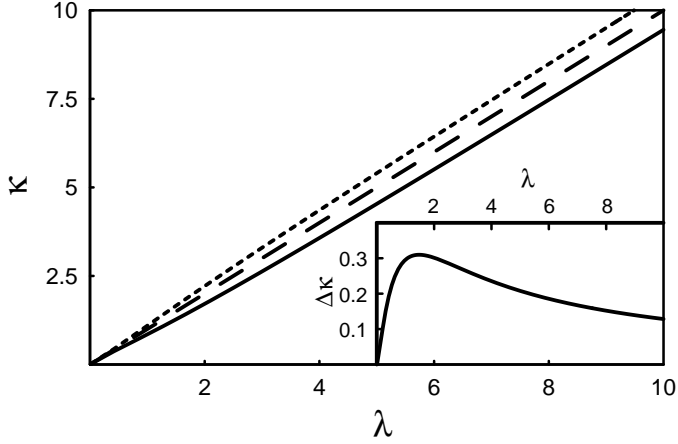}
\caption{Equilibrium condensate anisotropy
parameter $\kappa=\xi_r/\xi_z$ versus the trap
anisotropy parameter
$\lambda=\omega_z/\omega_r$. $\kappa$ is
calculated in the Thomas-Fermi regime with
$\varepsilon_{\mathrm{dd}}=0.29$. The three
curves correspond to three different values of
$\varphi$: the curve for $\varphi=0$ is solid,
for $\varphi=54.7^o$ is dashed and for
$\varphi=90^o$ it is dotted. The inset shows
$\Delta\kappa=(\kappa(\varphi=90^o)-\kappa(\varphi=0^o))/
\kappa(\varphi=57^o)$ versus the trap
anisotropy parameter $\lambda$.} \label{fig1}
\end{figure}

\section{Condensate expansion}

The expansion of the condensate (as well as
its equilibrium configuration) can be studied
using a time-dependent mean-field variation
approach similar to Ref. \cite{yi00,castin96}
(see also \cite{giova02b}) by assuming a
gaussian wave function as an ansatz
\begin{eqnarray}
\Psi(x,y,z,t)=\left(\frac{\pi^{-3/2} N}{\xi_x
\xi_y
\xi_z}\right)^{1/2}\prod_{\eta=x,y,z}e^{-\frac{\eta^2}{2
\xi^2_\eta} +i\eta^2\beta_\eta},\label{gausan}
\end{eqnarray}
where $\xi_\eta$ (width) and $\beta_\eta$ are
time dependent variational parameters. It is
easy to derive the following set of equation
for the condensate widths
\begin{eqnarray}
\ddot{\xi_\eta}&=&-\frac 2 M
{\partial\over\partial
\xi_\eta}\frac{H_{\mathrm{tot}}}{N}\left[
\xi_x,\xi_y,\xi_z\right] \label{var}
\end{eqnarray}
($\beta_\eta={M\dot{\xi_\eta}\over 2\hbar
\xi_\eta}$). The various contributions to the
total mean-field energy functional
$H_{\mathrm{tot}}$ evaluated on the gaussian
wave function are: (a) the kinetic energy in
the Thomas-Fermi limit $H_{\mathrm{kin}}= (N
M/4) \left(\dot{\xi}_x^2+ \dot{\xi}_y^2+
\dot{\xi}_z^2\right)$ (where we neglect here
the contribution of the zero-point
fluctuations $(N \hbar^2/4 M)
\left(1/\xi_x^2+1/\xi_y^2+1/\xi_z^2\right)$);
(b) the potential energy in the harmonic trap
$H_{\mathrm{ho}}=(N m/4)
\left(\omega_r^2\xi_x^2+\omega_r^2\xi_y^2 +
\omega_z^2\xi_z^2\right)$; (c) the mean-field
energy due to s-wave scattering
\begin{eqnarray}
H_{s}&=&{ N^2 \hbar^2 a \over m \sqrt{2\pi}}
{1\over \xi_r^2 \xi_z}\;;
\end{eqnarray}
(d) the mean-field magnetic dipole-dipole
energy, which for a cylindrically symmetric
condensate is given by
\begin{eqnarray}
H_{dd}&=&-\frac{N^2 \mu_0 \mu^2}{12
\pi\sqrt{2\pi}}{ f(\xi_r/\xi_z)\over \xi_r^2
\xi_z}
\label{hintvar}\\
f(\kappa)&=&\frac{1+2\kappa^2}{1-\kappa^2}-\frac{3\kappa^2
{\tanh}^{-1}\sqrt{1-\kappa^2}}{\left(1-\kappa^2\right)^{3/2}}
\end{eqnarray}
Here, we have introduced the radial width
$\xi_r=\xi_x=\xi_y$. The argument of $f$ is
the condensate aspect ratio
$\kappa=\xi_r/\xi_z$.

\begin{figure}
    \includegraphics[width=0.7\columnwidth]{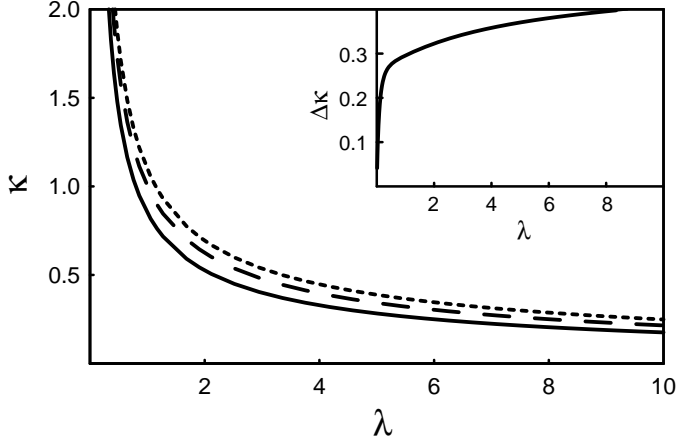}
\caption{Condensate anisotropy parameter
$\kappa$ after expansion versus the trap
anisotropy parameter $\lambda$. $\kappa$ is
calculated in the Thomas-Fermi regime with
$\varepsilon_{\mathrm{dd}}=0.29$. The three
curves correspond to three different values of
$\varphi$: the curve for $\varphi=0$ is solid,
for $\varphi=54.7^o$ is dashed and for
$\varphi=90^o$ it is dotted. The inset shows
$(\kappa(\varphi=90^o)-\kappa(\varphi=0^o))/
\kappa(\varphi=57^o)$ versus the trap
anisotropy parameter $\lambda$.
After expansion, the observability is improved
in the "pancake" geometry (large $\lambda$).}
\label{fig2}
\end{figure}

The anisotropy of the static dipole-dipole interaction is directly
manifest in the modification of the shape of the condensate. In
the Thomas-Fermi limit the ratio
\begin{eqnarray}
\kappa = \frac{\langle
x^2\rangle^{1/2}}{\langle
z^2\rangle^{1/2}}=\frac{\xi_r}{\xi_z}
\end{eqnarray}
of the radial $ \langle x^2\rangle^{1/2}$ and
longitudinal width $\langle z^2\rangle^{1/2}$
of the condensate is independent of the number
of atoms, because the dipolar energy has the
same scaling dependence as the s-wave
scattering energy. In particular $\kappa$
varies as function of the  "knob" angle
$\varphi$. The anisotropy parameter $\kappa$
of the condensate can be studied for an
equilibrium situation, in which the condensate
is harmonically trapped, or after a free
expansion. Generally, the condensate
anisotropy $\kappa$ is a function of the trap
anisotropy parameter
\begin{eqnarray}
\lambda = \omega_z/\omega_r \;.
\end{eqnarray}
In the Thomas-Fermi limit $\kappa$ is a
solution of the following equation
\begin{eqnarray}
\lambda^2 = \kappa^2 \frac
{1-\varepsilon_{\mathrm{dd}}
f(\kappa)-\varepsilon_{\mathrm{dd}}
\kappa\frac{\partial f}{\partial
\kappa}(\kappa)} {1-\varepsilon_{\mathrm{dd}}
f(\kappa)+\frac12 \varepsilon_{\mathrm{dd}}
\kappa\frac{\partial f}{\partial
\kappa}(\kappa)} \label{response}\;.
\end{eqnarray}
It is straightforward to see the standard
$\kappa=\lambda$ result for a trapped
condensate if dipolar forces are absent. When
$\varepsilon_{\mathrm{dd}} \ll 1$ we can
evaluate analytically the variation of the
aspect ratio $\kappa$ and the variation of the
"condensate volume" $V = (\langle x^2\rangle
\langle y^2\rangle \langle z^2\rangle)^{1/2}$
\begin{eqnarray}
\frac {\delta \kappa}{\kappa} &=&  \frac34
\varepsilon_{\mathrm{dd}} \, \lambda\,
\frac{\partial
f}{\partial \kappa}(\lambda)  \label{linearresponse}\\
\frac {\delta V}{V} &=& -
\varepsilon_{\mathrm{dd}} \left( \frac35
f(\lambda) + \frac 9{20} \lambda
\frac{\partial f}{\partial \kappa}(\lambda)
\right) \label{linearresponse2}\;.
\end{eqnarray}
The condensate's aspect ratio $\kappa$ is a
more universal parameter than the overall size
$V$ because it depends only on the trap
geometry and not on the number of atoms. In
particular the variation of $\kappa$ is easy
to measure because the value of $\kappa$ in
absence of dipolar forces is well know (in the
Thomas-Fermi limit, it is given by the trap
anisotropy $\lambda$; therefore $\delta
\kappa=\kappa-\lambda$). Moreover it is
possible to change the tuning angle $\varphi$
and measure the difference
$\kappa(\varphi=90^o)-\kappa(\varphi=0^o)$. In
contrast, the variation of the condensate
volume $V$ is more difficult to measure
because it requires the knowledge of the
number of atoms, a quantity that is usually
difficult to determine with sufficient
accuracy. From Eq. (\ref{linearresponse}) it
is possible to see that $\kappa$ is not
sensitive to the dipolar forces for high and
in particular for small values of $\lambda$
because the product $\kappa\frac{\partial
f}{\partial \kappa}(\kappa)$ goes to zero in
the two limits $\kappa\rightarrow 0$ and
$\kappa\rightarrow \infty$. The maximum linear
response of $\kappa$ to
$\varepsilon_{\mathrm{dd}}$, i.e. $\delta
\kappa \sim .64\, \varepsilon_{\mathrm{dd}} $,
is obtained for $\lambda\approx 2$.

Figure 1 shows the equilibrium anisotropy
parameter $\kappa$ of the condensate versus
the trap anisotropy parameter $\lambda$ in the
Thomas-Fermi regime, calculated for a chromium
condensate where a s-wave scattering length of
$a=2.8$ nm is assumed
($\varepsilon_{\mathrm{dd}}=0.29$). The effect
of the dipole interaction is to reduce
$\kappa$ for $\varphi=0^o$ or to increase
$\kappa$ for $\varphi=90^o$ with respect to
the equilibrium value of $\kappa$ in absence
of dipolar forces (see Figure 1). Note that by
varying $\varphi$, the total variation of
$\kappa$ can be increased. Thus, the influence
of the dipolar force becomes more visible.

\begin{figure}
    \includegraphics[width=0.7\columnwidth]{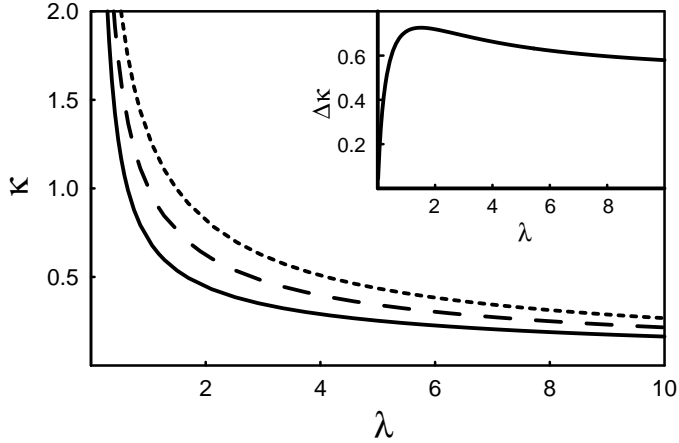}
    \caption{Condensate anisotropy parameter
$\kappa$ after a free expansion with
'inverted' dipolar coupling. When the
condensate is released, the angle $\varphi$ is
switched from $\varphi=0$ to $\varphi=90^o$.
$\kappa$ is calculated in the Thomas-Fermi
regime with $\varepsilon=0.29$ as function of
the trap anisotropy parameter $\lambda$. The
three curves correspond to three different
values of $\varphi$: the curve for $\varphi=0$
is solid, for $\varphi=54.7^o$ is dashed and
for $\varphi=90^o$ it is dotted. The inset
shows
$(\kappa(\varphi=90^o)-\kappa(\varphi=0^o))/
\kappa(\varphi=57^o)$ versus the trap
anisotropy parameter $\lambda$.}\label{fig3}
\end{figure}

Now we consider the expansion of the
condensate. Like for the static properties,
the evolution of the condensate anisotropy
parameter is independent of the number of
atoms in the Thomas-Fermi limit. Figure 2
shows the asymptotic values of the anisotropy
parameter $\kappa(t=\infty)$ after expansion.
In a finite time experiment these values are
approximated by the ratios of the velocities
$\dot{\xi}_r/\dot{\xi}_z$.

Similar to the equilibrium case the effect of
a (weak) dipole interaction is to reduce
$\kappa$ for $\varphi=0^o$ (or to increase
$\kappa$ for $\varphi=90^o$) with respect to
the values $\kappa(t=\infty)$ in absence of
dipolar forces (see Figure 2). In the absence
of dipolar forces $\kappa(t=\infty)$ is always
inverted after expansion.

According to the results shown in Figure 2, a
pancake trap will enhance the variation of
$\kappa$ with $\varphi$. Assuming $a=2.8$nm,
the variation of $\kappa$ is $40\%$. If
$\varepsilon_{\mathrm{dd}}$ is in reality
smaller than our assumption one should think
of some scheme to amplify the effect. A simple
scheme to do that is the following (see Fig.
3): when the condensate starts expanding, the
sign of the effective dipolar coupling should
be changed by changing the angle $\varphi$
from $0^o$ to $90^o$. With this simple
technique the observability is further
increased.

The particular case of a dipolar condensate
initially trapped in an isotropic trap
($\lambda=1$) deserves some consideration.
When the condensate is released its
anisotropy, i.e. its shape (cigar deformed for
$\varphi=0^o$ and pancake deformed for
$\varphi=90^o$), is conserved during the time
evolution. Indeed the set of equations
(\ref{var}) with a general time-dependent
isotropic harmonic potential frequency
$\omega_0(t)$ admits a solution of the form
$\xi_\eta(t)=\xi(t)\xi_\eta^0$ with
$\eta=x,y,z$ (where $\xi_\eta^0$ for instance
are stationary solution of (\ref{var}) with
$\omega_0(t)=\omega_0$). Similarly the
frequency of the monopole mode in an isotropic
trap is given by $\sqrt{5}\omega_0$, where
$\omega_0$ is the trap frequency,
irrespectively of the dipolar forces.

\section{Conclusions}

We have theoretically investigate the static
properties as well the expansion of a dipolar
condensate in the Thomas-Fermi regime. We
found that the pancake geometry is a more
suitable trap geometry to observe dipolar
effects, since they are enhanced during
expansion.

\emph {Note added in proof} --- We are aware
of one similar work on the expansion of a
dipolar condensate \cite{yiyou}.

\ack Funding was provided by the RTN network
"cold quantum gases" under the contract number
HPRN-CT-2000-00125, and the Deutsche
Forschungsgemeinschaft. We thank Kazimierz
Rz\c{a}\.{z}ewski for a critical reading of
this manuscript.


\begin{thebibliography}{10}
\expandafter\ifx\csname url\endcsname\relax
  \def\url#1{\texttt{#1}}\fi
\expandafter\ifx\csname
urlprefix\endcsname\relax\def\urlprefix{URL
}\fi


\bibitem{stringari} F. Dalfovo {\em et al.}, Rev. Mod. Phys. \textbf{ 71}, 463 (1999).

\bibitem{yi00} S. Yi and L. You, Phys. Rev. A \textbf{ 61}, 041604
(2000); S. Yi and L. You, Phys. Rev. A
\textbf{ 63}, 053607 (2001).

\bibitem{goral00} K. Goral, K. Rzazewski, and T. Pfau,
Phys. Rev. A  \textbf{ 61}, 051601 (2000);
J.-P. Martikainen, Matt Mackie, and K.-A.
Suominen, Phys. Rev. A  \textbf{ 64}, 037601
(2001).

\bibitem{santos00} L. Santos {\em et al.}, Phys. Rev. Lett.
\textbf{ 85}, 1791 (2000); L. Santos {\em et
al.}, Phys. Rev. Lett. \textbf{ 88}, 139904
(2002).

\bibitem{baranov} M.A. Baranov {\em et al.}, cond-mat/0109437.

\bibitem{goral} K. Goral, L. Santos and M. Lewenstein, Phys.
Rev. Lett. \textbf{ 88}, 170406 (2002).

\bibitem{meystre01}
H. Pu, W. Zhang, and P. Meystre, Phys. Rev.
Lett. \textbf{ 87}, 140405 (2001); W. Zhang,
H. Pu, C. Search, and P. Meystre, Phys. Rev.
Lett. \textbf{ 88}, 060401 (2002).

\bibitem{deMille} D. DeMille, Phys. Rev. Lett. \textbf{ 88}, 067901
(2002).

\bibitem{giova02} D. O'Dell,  et al., Phys. Rev.
Lett. \textbf{ 84}, 5687 (2000); S.
Giovanazzi, D. O'Dell, and G. Kurizki, Phys.
Rev. A \textbf{ 63}, 031603 (2001); S.
Giovanazzi {\em et al.}, Europhys. Lett.
\textbf{ 56}, 1 (2001); S. Giovanazzi, D.
O'Dell, and G. Kurizki, Phys. Rev. Lett.
\textbf{ 88}, 130402 (2002).

\bibitem{santos02} M. Baranov {\em et al.}, cond-mat/0201100.

\bibitem{giova02b} S. Giovanazzi, A. G\"{o}rlitz, and T. Pfau, Phys. Rev. Lett.
\textbf{ 89}, 130401 (2002).

\bibitem{weinstein98} J.D. Weinstein, et al., Nature (London) \textbf{ 395}, 148 (1998).

\bibitem{weinstein98b} J. Weinstein, et al., Phys. Rev. A \textbf{ 57}, 3173 (1998).

\bibitem{goral2} K. Goral, and L. Santos, cond-mat/0203542; S. Yi
and L. You, cond-mat/0111256.

\bibitem{castin96} Y. Castin, and R. Dum, Phys. Rev. Lett.
\textbf{ 77}, 5315 (1996).

\bibitem{yiyou}  S. Yi
and L. You, cond-mat/0210677.


\end{thebibliography}
\end{document}